\documentclass[12pt]{article}

\begin{document}

\title{\bf Energy Distribution associated with Static Axisymmetric Solutions}

\author{M. Sharif \thanks{e-mail: msharif@math.pu.edu.pk} and Tasnim Fatima\\
Department of Mathematics, University of the Punjab,\\
Quaid-e-Azam Campus, Lahore-54590, Pakistan.}

\date{}

\maketitle

\begin{abstract}
This paper has been addressed to a very old but burning problem of
energy in General Relativity. We evaluate energy and momentum
densities for the static and axisymmetric solutions. This
specializes to two metrics, i.e., Erez-Rosen and the gamma
metrics, belonging to the Weyl class. We apply four well-known
prescriptions of Einstein, Landau-Lifshitz, Papaterou and
M$\ddot{o}$ller to compute energy-momentum density components. We
obtain that these prescriptions do not provide similar energy
density, however momentum becomes constant in each case. The
results can be matched under particular boundary conditions.
\end{abstract}


\section{Introduction}

The problem of energy-momentum of a gravitational field has always
been an attractive issue in the theory of General Relativity (GR).
The notion of energy-momentum for asymptotically flat spacetime is
unanimously accepted. Serious difficulties in connection with its
notion arise in GR. However, for gravitational fields, this can be
made locally vanish. Thus one is always able to find the frame in
which the energy-momentum of gravitational field is zero, while in
other frames it is not true. Noether's theorem and translation
invariance lead to the canonical energy-momentum density tensor,
$T_a^b$, which is conserved.
\begin{equation}
T^b_{a;b}=0,\quad (a,b=0,1,2,3).
\end{equation}
In order to obtain a meaningful expression for energy-momentum, a
large number of definitions for the gravitation energy-momentum in
GR have been proposed. The first attempt was made by Einstein who
suggested an expression for energy-momentum density [1]. After
this, many physicists including Landau-Lifshitz [2], Papapetrou
[3], Tolman [4], Bergman [5] and Weinburg [6] had proposed
different expressions for energy-momentum distribution. These
definitions of energy-momentum complexes give meaningful results
when calculations are performed in Cartesian coordinates. However,
the expressions given by M$\ddot{o}$ller [7,8] and Komar [9] allow
one to compute the energy-momentum densities in any spatial
coordinate system. An alternate concept of energy, called
quasi-local energy, does not restrict one to use particular
coordinate system. A large number of definitions of quasi-local
masses have been proposed by Penrose [10] and many others [11,12].
Chang et al. [13] showed that every energy-momentum complex can be
associated with distinct boundary term which gives the quasi-local
energy-momentum.

There is a controversy with the importance of non-tensorial
energy-momentum complexes whose physical interpretation has been a
problem for the scientists. There is a uncertainity that different
energy-momentum complexes would give different results for a given
spacetime. Many researchers considered different energy-momentum
complexes and obtained encouraging results. Virbhadra et al.
[14-18] investigated several examples of the spacetimes and showed
that different energy-momentum complexes could provide exactly the
same results for a given spacetime. They also evaluated the
energy-momentum distribution for asymptotically non-flat
spacetimes and found the contradiction to the previous results
obtained for asymptotically flat spacetimes. Xulu [19,20]
evaluated energy-momentum distribution using the M$\ddot{o}$ller
definition for the most general non-static spherically symmetric
metric. He found that the result is different in general from
those obtained using Einstein's prescription. Aguirregabiria et
al. [21] proved the consistency of the results obtained by using
the different energy-momentum complexes for any Kerr-Schild class
metric.

On contrary, one of the authors (MS) considered the class of
gravitational waves, G$\ddot{o}$del universe and homogeneous
G$\ddot{o}$del-type metrics [22-24] and used the four definitions
of the energy-momentum complexes. He concluded that the four
prescriptions differ in general for these spacetimes. Ragab
[25,26] obtained contradictory results for G$\ddot{o}$del-type
metrics and Curzon metric which is a special solution of the Weyl
metrics. Patashnick [27] showed that different prescriptions give
mutually contradictory results for a regular MMaS-class black
hole. In recent papers, we extended this procedure to the non-null
Einstein-Maxwell solutions, electromagnetic generalization of
G$\ddot{o}$del solution, singularity-free cosmological model and
Weyl metrics [28-30]. We applied four definitions and concluded
that none of the definitions provide consistent results for these
models. This paper continues the study of investigation of the
energy-momentum distribution for the family of Weyl metrics by
using the four prescriptions of the energy-momentum complexes. In
particular, we would explore energy-momentum for the Erez-Rosen
and gamma metrics.

The paper has been distributed as follows. In the next section, we
shall describe the Weyl metrics and its two family members
Erez-Rosen and gamma metrics. Section 3 is devoted to the
evaluation of energy-momentum densities for the Erez-Rosen metric
by using the prescriptions of Einstein, Landau-Lifshitz,
Papapetrou and M$\ddot{o}$ller. In section 4, we shall calculate
energy-momentum density components for the gamma metric. The last
section contains discussion and summary of the results.

\section{The Weyl Metrics}

Static axisymmetric solutions to the Einstein field equations are
given by the Weyl metric [31,32]
\begin{equation}
ds^2=e^{2\psi}dt^2-e^{-2\psi}[e^{2\gamma}(d\rho^2+dz^2)
+\rho^2d\phi^2]
\end{equation}
in the cylindrical coordinates $(\rho,~\phi,~z)$. Here $\psi$ and
$\gamma$ are functions of coordinates $\rho$ and $z$. The metric
functions satisfy the following differential equations
\begin{eqnarray}
\psi_{\rho\rho}+\frac{1}{\rho}\psi_{\rho}+\psi_{zz}=0,\\
\gamma_{\rho}=\rho(\psi^2_{\rho}-\psi^2_{z}),\quad
\gamma_{z}=2\rho\psi_{\rho}\psi_{z}.
\end{eqnarray}
It is obvious that Eq.(3) represents the Laplace equation for
$\psi$. Its general solution, yielding an asymptotically flat
behaviour, will be
\begin{equation}
\psi=\sum^\infty_{n=0}\frac{a_n}{r^{n+1}}P_n(\cos\theta),
\end{equation}
where $r=\sqrt{\rho^2+z^2},~\cos\theta=z/r$ are Weyl spherical
coordinates and $P_n(\cos\theta)$ are Legendre Polynomials. The
coefficients $a_n$ are arbitrary real constants which are called
{\it Weyl moments}. It is mentioned here that if we take
\begin{eqnarray}
\psi=-\frac{m}{r},\quad\gamma=-\frac{m^2\rho^2}{2r^4},\quad
r=\sqrt{\rho^2+z^2}
\end{eqnarray}
then the Weyl metric reduces to special solution of Curzon metric
[33]. There are more interesting members of the Weyl family,
namely the Erez-Rosen and the gamma metric whose properties have
been extensively studied in the literature [32,34].

The Erez-Rosen metric [32] is defined by considering the special
value of the metric function
\begin{equation}
2\psi=ln(\frac{x-1}{x+1})+q_2(3y^2-1)[\frac{1}{4}(3x^2-1)
ln(\frac{x-1}{x+1})+\frac{3}{2}x],
\end{equation}
where $q_2$ is a constant.

\section{Energy and Momentum for the Erez-Rosen Metric}

In this section, we shall evaluate the energy and momentum density
components for the Erez-Rosen metric by using different
prescriptions. To obtain meaningful results in the prescriptions
of Einstein, Ladau-Lifshitz's and Papapetrou, it is required to
transform the metric in Cartesian coordinates. This can be done by
using the transformation equations
\begin{equation}
x=\rho cos\theta,\quad y=\rho sin\theta.
\end{equation}
The resulting metric in these coordinates will become
\begin{equation}
ds^2=e^{2\psi}dt^2-\frac{e^{2(\gamma-\psi)}}{\rho^2}(xdx+ydy)^2\nonumber\\
-\frac{e^{-2\psi}}{\rho^2}(xdy-ydx)^2-e^{2(\gamma-\psi)}dz^2.
\end{equation}

\subsection{Energy and Momentum in Einstein's Prescription}

The energy-momentum complex of Einstein [1] is given by
\begin{equation}
\Theta^b_a= \frac{1}{16 \pi}H^{bc}_{a,c},
\end{equation}
where
\begin{equation}
H^{bc}_a=\frac{g_{ad}}{\sqrt{-g}}[-g(g^{bd}g^{ce}
-g^{be}g^{cd})]_{,e},\quad a,b,c,d,e = 0,1,2,3.
\end{equation}
Here $\Theta^0_{0}$ is the energy density, $\Theta^i_{0}~
(i=1,2,3)$ are the momentum density components and $\Theta^0_{i}$
are the energy current density components. The Einstein
energy-momentum satisfies the local conservation laws
\begin{equation}
\frac{\partial \Theta^b_a}{\partial x^{b}}=0.
\end{equation}
The required components of $H_a^{bc}$  are the following
\begin{eqnarray}
H^{01}_{0}&=&\frac{4y}{\rho^2}e^{2\gamma}(y\psi_{,x}-x\psi_{,y})
+\frac{4x}{\rho^2}(x\psi_{,x}+y\psi_{,y})\nonumber\\
&-&\frac{x}{\rho^2}-2x\psi^2_{,\rho}+\frac{x}{\rho^2}e^{2\gamma},\\
H^{02}_{0}&=&\frac{4x}{\rho^2}e^{2\gamma}(x\psi_{,y}-y\psi_{,x})
+\frac{4y}{\rho^2}(x\psi_{,x}+y\psi_{,y})\nonumber\\
&-&\frac{y}{\rho^2}-2y\psi^2_{,\rho}+\frac{y}{\rho^2}e^{2\gamma}.
\end{eqnarray}
Using Eqs.(13)-(14) in Eq.(10), we obtain the energy and momentum
densities in Einstein's prescription
\begin{eqnarray}
\Theta^0_{0}&=&\frac{1}{8\pi
\rho^2}[e^{2\gamma}\{\rho^2\psi^2_{,\rho}+2(x^2\psi_{,yy}+y^2\psi_{xx}
-x\psi_{,x}-y\psi_{,y})\}\nonumber\\
&+&2\{x^2\psi_{,xx}+y^2\psi_{,yy}+x\psi_{,x}+y\psi_{,y}
-\rho^2\psi_{,\rho}(\psi_{,\rho}+\rho \psi_{,\rho\rho})\}].
\end{eqnarray}
All the momentum density components turn out to be zero and hence
momentum becomes constant.

\subsection{Energy and Momentum in Landau-Lifshitz's Prescription}

The Landau-Lifshitz [2] energy-momentum complex can be written as
\begin{equation}
L^{ab}= \frac{1}{16 \pi}\ell^{acbd}_{,cd},
\end{equation}
where
\begin{equation}
\ell^{acbd}= -g(g^{ab}g^{cd}-g^{ad}g^{cb}).
\end{equation}
$L^{ab}$ is symmetric with respect to its indices. $L^{00}$ is the
energy density and $L^{0i}$ are the momentum (energy current)
density components. $\ell^{abcd}$ has symmetries of the Riemann
curvature tensor. The local conservation laws for Landau-Lifshitz
energy-momentum complex turn out to be
\begin{equation}
\frac{\partial L^{ab}}{\partial x^{b}}=0.
\end{equation}
The required non-vanishing components of $\ell^{acbd}$ are
\begin{eqnarray}
\ell^{0101}&=&-\frac{y^2}{\rho^2}e^{4\gamma-4\psi}
-\frac{x^2}{\rho^2}e^{2\gamma-4\psi},\\
\ell^{0202}&=&-\frac{x^2}{\rho^2}e^{4\gamma-4\psi}
-\frac{y^2}{\rho^2}e^{2\gamma-4\psi},\\
\ell^{0102}&=&\frac{xy}{\rho^2}e^{4\gamma-4\psi}
-\frac{xy}{\rho^2}e^{2\psi-4\gamma}.
\end{eqnarray}
Using Eqs.(19)-(21) in Eq.(16), we get
\begin{eqnarray}
L^{00}&=&\frac{e^{2\gamma-4\psi}}{8\pi
\rho^2}[e^{2\gamma}\{2\rho^2\psi^2_{,\rho}-8(y^2\psi^2_{,x}
+x^2\psi^2_{,y})+2(x^2\psi_{,xx}+y^2\psi_{,yy}\nonumber\\
&-&x\psi_{,x}-y\psi_{,y})+16xy\psi_{,x}\psi_{,y}-4xy\psi_{,xy}\}\nonumber\\
&-&\rho^2\psi_{,\rho}(3\psi_{,\rho} +2\rho^2\psi^3_{,\rho}+2\rho
\psi_{,\rho\rho})-8\rho^2\psi^2_{,\rho}(x\psi_{,x}+y\psi_{,y})\nonumber\\
&-&8(x^2\psi^2_{,x}+y^2\psi^2_{,y})+2(x^2\psi_{,xx}+y^2\psi_{,yy}\nonumber\\
&+&x\psi_{,x}+y\psi_{,y})-16xy\psi_{,x}\psi_{,y}+4xy\psi_{,xy}].
\end{eqnarray}
The momentum density vanishes and hence momentum becomes constant.

\subsection{Energy and Momentum in Papapetrou's Prescription}

We can write the prescription of Papapetrou [3] energy-momentum
distribution in the following way
\begin{equation}
\Omega^{ab}=\frac{1}{16\pi}N^{abcd}_{,cd},
\end{equation}
where
\begin{equation}
N^{abcd}=\sqrt{-g}(g^{ab}\eta^{cd}-g^{ac}\eta^{bd}
+g^{cd}\eta^{ab}-g^{bd}\eta^{ac}),
\end{equation}
and $\eta^{ab}$ is the Minkowski spacetime. It follows that the
energy-momentum complex satisfies the following local conservation
laws
\begin{equation}
\frac{\partial \Omega^{ab}}{\partial x^b}=0.
\end{equation}
$\Omega^{00}$ and $\Omega^{0i}$ represent the energy and momentum
(energy current) density components respectively.

The required components of $N^{abcd}$ are
\begin{eqnarray}
N^{0011}&=&-\frac{y^2}{\rho^2}e^{2\gamma}-\frac{x^2}{\rho^2}
-e^{2\gamma-4\psi},\\
N^{0022}&=&-\frac{x^2}{\rho^2}e^{2\gamma}-\frac{y^2}{\rho^2}
-e^{2\gamma-4\psi},\\
N^{0012}&=&-\frac{xy}{\rho^2}e^{2\gamma}-\frac{xy}{\rho^2}.
\end{eqnarray}
Substituting Eqs.(26)-(28) in Eq.(23), we obtain the following
energy density
\begin{eqnarray}
\Omega^{00}&=&\frac{e^{2\gamma}}{8\pi}[\psi^2_{,\rho}
-e^{-4\psi}\{\psi^2_{,\rho}+2\rho^2\psi^4_{,\rho} +2\rho
\psi_{,\rho}\psi_{,\rho\rho}\nonumber\\
&-&8\psi^2_{,\rho}(x\psi_{,x}+y\psi_{,y})+
8(\psi^2_{,x}+\psi^2_{,y})-2(\psi_{,xx}+\psi_{,yy})\}].
\end{eqnarray}
The momentum density vanishes.

\subsection{Energy and Momentum in M\"{o}ller's Prescription}

The energy-momentum density components in M\"{o}ller's
prescription [7,8] are given as
\begin{equation}
M^b_a= \frac{1}{8\pi}K^{bc}_{a,c},
\end{equation}
where
\begin{equation}
K_a^{bc}= \sqrt{-g}(g_{ad,e}-g_{ae,d})g^{be}g^{cd}.
\end{equation}
Here $K^{bc}_{ a}$ is symmetric with respect to the indices.
$M^0_{0}$ is the energy density, $M^i_{0}$ are momentum density
components, and $M^0_{i}$ are the components of energy current
density. The M\"{o}ller energy-momentum satisfies the following
local conservation laws
\begin{equation}
\frac{\partial M^b_a}{\partial x^b}=0.
\end{equation}
Notice that M\"{o}ller's energy-momentum complex is independent of
coordinates.

The components of $K^{bc}_a$ for Erez-Rosen metric is the
following
\begin{eqnarray}
K^{01}_0&=&2\rho \psi_{,\rho}.
\end{eqnarray}
Substitute Eq.(33) in Eq.(30), we obtain
\begin{eqnarray}
M^0_0&=&\frac{1}{4\pi}[\psi_{,\rho}+\rho\psi_{,\rho\rho}].
\end{eqnarray}
Again, we get momentum constant.

The partial derivatives of the function $\psi$ are given by
\begin{eqnarray}
\psi_{,x}&=&\frac{1}{x^2-1}+\frac{q_2}{4}(3y^2-1)[3x
ln(\frac{x-1}{x+1})+\frac{3x^2-1}{x^2-1}+3],\\
\psi_{,y}&=&\frac{3yq_2}{4}[(3x^2-1)
ln(\frac{x-1}{x+1})+6x],\\
\psi_{,xx}&=&\frac{-2x}{(x^2-1)^2}+\frac{q_2}{4}(3y^2-1)[3
ln(\frac{x-1}{x+1})+2x\frac{3x^2-5}{(x^2-1)^2}],\\
\psi_{,yy}&=&\frac{3q_2}{4}[(3x^2-1)
ln(\frac{x-1}{x+1})+6x],\\
\psi_{,xy}&=& U_{,yx}=\frac{3yq_2}{4}[3x
ln(\frac{x-1}{x+1})+2\frac{3x^2-2}{x^2-1}],\\
\psi_{,\rho}&=& \frac{\rho}{x(x^2-1)}+\frac{\rho
q_2}{4x}[3x(3\rho^2-2)
ln(\frac{x-1}{x+1})\nonumber\\
&+&2\frac{(3x^2-1)(3y^2-1)}{x^2-1}+18x^2],
\end{eqnarray}
\newpage
\begin{eqnarray}
\psi_{,\rho\rho}&=&\frac{1}{x(x^2-1)}-\frac{2\rho^2}{x(x^2-1)^2}
+\frac{q_2}{4x^2}(3y^2-1)[3(\rho^2+x^2)
ln(\frac{x-1}{x+1})\nonumber\\
&+&\frac{2x}{x^2-1}(3x^2-2+\frac{\rho^2(3x^2-5)}{x^2-1})]
+\frac{3\rho
q_2}{4}(1+\frac{\rho}{y^2})\nonumber\\
&\times&[(3x^2-1)ln(\frac{x-1}{x+1})+6x]+\frac{3\rho^2q_2}{x}[3
ln(\frac{x-1}{x+1})+2\frac{3x^2-1}{x^2-1}].
\end{eqnarray}

\section{Energy and Momentum for the Gamma Metric}

A static and asymptotically flat exact solution to the Einstein
vacuum equations is known as the gamma metric. This is given by
the metric [34]
\begin{equation}
ds^2=(1-\frac{2m}{r})^{\gamma}dt^2-(1-\frac{2m}{r})^{-\gamma}
[(\frac{\Delta}{\Sigma})^{\gamma^2-1}dr^2+\frac{\Delta^{\gamma^2}}
{\Sigma^{\gamma^2-1}}d\theta^2+\Delta\sin^2\theta d\phi^2],
\end{equation}
where
\begin{eqnarray}
\Delta &=& r^2-2mr,\\
\Sigma &=& r^2-2mr+m^2sin^2\theta,
\end{eqnarray}
$m$ and $\gamma$ are constant parameters. $m=0$ or $\gamma=0$
gives the flat spacetime. For $|\gamma|=1$ the metric is
spherically symmetric and for $|\gamma|\neq1$, it is axially
symmetric. $\gamma=1$ gives the Schwarzschild spacetime in the
Schwarzschild coordinates. $\gamma=-1$ gives the Schwarzschild
spacetime with negative mass, as putting $m=-M(m>0)$ and carrying
out a non-singular coordinate transformation $(r\rightarrow
R=r+2M)$ one gets the Schwarzschild spacetime (with positive mass)
in the Schwarzschild coordinates $(t,R,\theta,\Phi)$.

In order to have meaningful results in the prescriptions of
Einstein, Landau-Lifshitz and Papapetrou, it is necessary to
transform the metric in Cartesian coordinates. We transform this
metric in Cartesian coordinates by using
\begin{equation}
x=rsin\theta\cos\phi,\quad y=rsin\theta\sin\phi,\quad
z=rcos\theta.
\end{equation}
The resulting metric in these coordinates will become
\begin{eqnarray}
ds^2&=&(1-\frac{2m}{r})^{\gamma}dt^2-(1-\frac{2m}{r})^{-\gamma}
[(\frac{\Delta}{\Sigma})^{\gamma^2-1}\frac{1}{r^2}
\{xdx+ydy+zdz\}^2\nonumber\\
&+&\frac{\Delta^{\gamma^2}}{\Sigma^{\gamma^2-1}}
\{\frac{xzdx+yzdy-(x^2+y^2)dz}{r^2\sqrt{x^2+y^2}}\}^2
+\frac{\Delta(xdy-ydx)^2}{r^2(x^2+y^2)}.
\end{eqnarray}
Now we calculate energy-momentum densities using the different
prescriptions given below.

\subsection{Energy and Momentum in Einstein's Prescription}

The required non-vanishing components of $H^{bc}_{a}$ are
\begin{eqnarray}
H^{01}_{0}&=&4\gamma m \frac{x}{r^3}+(\frac{\Delta}
{\Sigma})^{\gamma^2-1}\frac{x}{x^2+y^2}-(\gamma^2+1)
(1-\frac{m}{r})\frac{2x}{r^2}\nonumber\\
&+& (\gamma^2-1)(1-\frac{m}{r})\frac{2\Delta x}{\Sigma
r^2}+\frac{2\Delta x}{r^4}
+(\gamma^2-1)\frac{2m^2xz^2}{\Sigma r^4}\nonumber\\
&-&\frac{xz^2}{r^2(x^2+y^2)}+\frac{x}{r^2},\\
H^{02}_{0}&=&4\gamma m \frac{y}{r^3}+(\frac{\Delta}
{\Sigma})^{\gamma^2-1}\frac{y}{x^2+y^2}-(\gamma^2+1)
(1-\frac{m}{r})\frac{2y}{r^2}\nonumber\\
&+& (\gamma^2-1)(1-\frac{m}{r})\frac{2\Delta y}{\Sigma
r^2}+\frac{2\Delta y}{r^4}
+(\gamma^2-1)\frac{2m^2yz^2}{\Sigma r^4}\nonumber\\
&-&\frac{xyz^2}{r^2(x^2+y^2)}+\frac{y}{r^2},\\
H^{03}_{0}&=&4\gamma m
\frac{z}{r^3}-(\gamma^2+1)(1-\frac{m}{r})\frac{2z}{r^2}+
(\gamma^2-1)(1-\frac{m}{r})\frac{2\Delta z}{\Sigma
r^2}\nonumber\\&+&\frac{2\Delta z}{r^4}
-(\gamma^2-1)(x^2+y^2)\frac{2m^2z}{\Sigma r^4}+\frac{2z}{r^2}.
\end{eqnarray}
Using Eqs.(47)-(49) in Eq.(10), we obtain non-vanishing energy
density in Einstein's prescription given as
\begin{eqnarray}
\Theta^0_{0}&=&\frac{1}{8\pi
\Sigma^{\gamma^2}r^6}[(\gamma^2-1)\Sigma
\Delta^{\gamma^2-2}r^5(r-m)-(\gamma^2-1)\Delta^{\gamma^2-1}r^2\nonumber\\
&\times&\{r^4-mr^3+m^2r^2-m^2(x^2+y^2)\}
-(\gamma^2+1)\Sigma^{\gamma^2}r^4\nonumber\\
&+&2(\gamma^2-1)\Sigma^{\gamma^2-1} r^4(r-m)^2
+(\gamma^2-1)m\Delta\Sigma^{\gamma^2-1}r^2\nonumber\\
&-&(\gamma^2-1)\Delta^{\gamma^2-2}\Delta
r^4(r-m)^2+(\gamma^2-1)\Delta^{\gamma^2-1}\nonumber\\&\times&
\Delta r^5(r-m)+2\Sigma^{\gamma^2}r^3(r-m)-\Sigma^{\gamma^2}\Delta
r^2+\Sigma^{\gamma^2}r^4\nonumber\\
&+&3(\gamma^2-1)\Sigma^{\gamma^2}r^2m^2z^2
-(\gamma^2-1)\Sigma^{\gamma^2-1}r^4m^2\nonumber\\
&-&2(\gamma^2-1) \Sigma^{\gamma^2-2}m^4z^2(x^2+y^2)].
\end{eqnarray}
The momentum density components become zero and consequently
momentum is constant.

\subsection{Energy and Momentum in Landau-Lifshitz's Prescription}

The required non-vanishing components of $\ell^{acbd}$ are
\begin{eqnarray}
\ell^{0101}&=&-(1-\frac{2m}{r})^{-2\gamma}[\frac{y^2\Delta^{2\gamma^2-1}}
{r^2(x^2+y^2)\Sigma^{2(\gamma^2-1)}}
+\frac{x^2\Delta^{\gamma^2+1}}{r^6\Sigma^{\gamma^2-1}}\nonumber\\&
+&\frac{\Delta^{\gamma^2}x^2z^2}
{r^4(x^2+y^2)\Sigma^{\gamma^2-1}}],\\
\ell^{0202}&=&-(1-\frac{2m}{r})^{-2\gamma}[\frac{x^2\Delta^{2\gamma^2-1}}
{r^2(x^2+y^2)\Sigma^{2(\gamma^2-1)}}
+\frac{y^2\Delta^{\gamma^2+1}}{r^6\Sigma^{\gamma^2-1}}\nonumber\\&
+&\frac{\Delta^{\gamma^2}y^2z^2}
{r^4(x^2+y^2)\Sigma^{\gamma^2-1}}],\\
\ell^{0303}&=&-(1-\frac{2m}{r})^{-2\gamma}[\frac{z^2\Delta^{\gamma^2+1}}
{r^6\Sigma^{\gamma^2-1}}+\frac{(x^2+y^2)\Delta^{\gamma^2}}
{r^4\Sigma^{\gamma^2-1}}],\\
\ell^{0102}&=&(1-\frac{2m}{r})^{-2\gamma}[\frac{xy\Delta^{2\gamma^2-1}}
{r^2(x^2+y^2)\Sigma^{2(\gamma^2-1)}}
-\frac{xy\Delta^{\gamma^2+1}}{r^6\Sigma^{\gamma^2-1}}\nonumber\\&
-&\frac{\Delta^{\gamma^2}xyz^2}
{r^4(x^2+y^2)\Sigma^{\gamma^2-1}}],\\
\ell^{0103}&=&-(1-\frac{2m}{r})^{-2\gamma}[\frac{xz\Delta^{\gamma^2+1}}
{r^6\Sigma^{\gamma^2-1}}-\frac{xz\Delta^{\gamma^2}}
{r^4\Sigma^{\gamma^2-1}}],\\
\ell^{0203}&=&-(1-\frac{2m}{r})^{-2\gamma}[\frac{yz\Delta^{\gamma^2+1}}
{r^6\Sigma^{\gamma^2-1}}-\frac{yz\Delta^{\gamma^2}}
{r^4\Sigma^{\gamma^2-1}}].
\end{eqnarray}
When we substitute these values in Eq.(16), it follows that the
energy density remains non-zero while momentum density components
vanish. This is given as follows
\begin{eqnarray}
L^{00}&=&\frac{(1-\frac{2m}{r})^{-2\gamma}}{8\pi }[-\frac{4\gamma
m^2}{r^4}(\frac{\Delta}{\Sigma})^{\gamma^2-1}(2\gamma+1)+\frac{2\gamma
m}{r^7}\{-(\frac{\Delta}{\Sigma})^{2(\gamma^2-1)}r^4\nonumber\\
&+&4(\gamma^2+1)(1-\frac{m}{r})(\frac{\Delta}{\Sigma})^{\gamma^2-1}r^4
-4(\gamma^2-1)(1-\frac{m}{r})(\frac{\Delta}{\Sigma})^{\gamma^2}r^4\nonumber\\
&-&\frac{\Delta^{\gamma^2}}{\Sigma^{\gamma^2-1}}r^2-
(\frac{\Delta}{\Sigma})^{\gamma^2-1}r^4+4({\gamma^2-1})
\frac{\Delta^{\gamma^2-1}}{\Sigma^{\gamma^2}}m^2z^2(x^2+y^2)\}\nonumber\\
&+&\frac{1}{r^2}(2\gamma^2-1)(1-\frac{m}{r})(\frac{\Delta}
{\Sigma})^{2(\gamma^2-1)}-\frac{2}{r^2}(\gamma^2-1)(1-\frac{m}{r}\nonumber\\
&+&\frac{m^2}{r^2}-\frac{m^2(x^2+y^2)}{r^4})(\frac{\Delta}
{\Sigma})^{2\gamma^2-1}-\frac{\Delta^{2\gamma^2-1}}
{r^4\Sigma^{2(\gamma^2-1)}}\nonumber\\
&-&\frac{2\gamma^2}{r^2}(\gamma^2+1)(1-\frac{m}{r})^2(\frac{\Delta}
{\Sigma})^{\gamma^2-1}+\frac{4}{r^2}({\gamma^4-1})
(1-\frac{m}{r})^2(\frac{\Delta}{\Sigma})^{\gamma^2}\nonumber\\
&-&\frac{2}{r^2}({\gamma^2-1})(1-\frac{m}{r})^2(\frac{\Delta}
{\Sigma})^{\gamma^2+1}-2({\gamma^2-1})(1-\frac{m}{r}+\frac{m^2}{r^2}\nonumber\\
&-&\frac{m^2(x^2+y^2)}{r^4})\frac{\Delta^{\gamma^2+1}}{\Sigma^{\gamma^2}}-
(\gamma^2+1)\frac{m\Delta^{\gamma^2}}{r^5\Sigma^{\gamma^2}-1}\nonumber\\
&+&3(\gamma^2+1)\frac{\Delta^{\gamma^2}}{r^4\Sigma^{\gamma^2-1}}(1-\frac{m}{r})
+(\gamma^2-1)\frac{m\Delta^{\gamma^2+1}}{r^5\Sigma^{\gamma^2}}(1-\frac{2m}{r}\nonumber\\
&-& \frac{m^2(x^2+y^2)}{r^3}) -2(\gamma^2-1)(x^2+y^2)
\frac{m^2z^2\Delta^{\gamma^2+1}}{r^{10}\Sigma^{\gamma^2}}\nonumber\\&-&\frac{3\Delta^
{\gamma^2+1}}{r^6\Sigma^{\gamma^2-1}}+\gamma^2(\frac{\Delta}
{\Sigma})^{\gamma^2-1}(1-\frac{m}{r})(\frac{1}{r^2}+\frac{2z^2}{r^4})
-2\gamma^2(\gamma^2-1)\nonumber\\&\times&(x^2+y^2)\frac{m^4z^2\Delta^{\gamma^2}}
{r^8\Sigma^{\gamma^2+1}} +2(\gamma^2-1)(x^2+y^2)(\frac{\Delta}
{\Sigma})^{\gamma^2}\frac{m^2z^2}{r^8}\nonumber\\&-&(\gamma^2-1)(\frac{\Delta}
{\Sigma})^{\gamma^2}(1-\frac{m}{r}+\frac{m^2}{r^2})\frac{1}{r^2}.
\end{eqnarray}
As momentum density vanishes hence it is constant.

\subsection{Energy and Momentum in Papapetrou's Prescription}

The required non-vanishing components of $N^{abcd}$ are given by
\begin{eqnarray}
N^{0011}&=&-(\frac{\Delta}{\Sigma})^{\gamma^2-1}\frac{y^2}{x^2+y^2}
-\frac{\Delta x^2}{r^4}-\frac{x^2z^2}{r^2(x^2+y^2)}\nonumber\\
&-&(1-\frac{2m}{r})^{-2\gamma}
\frac{\Delta^{\gamma^2}}{r^2\Sigma^{\gamma^{2}-1}},\\
N^{0022}&=&-(\frac{\Delta}{\Sigma})^{\gamma^2-1}\frac{x^2}{x^2+y^2}
-\frac{\Delta y^2}{r^4}-\frac{y^2z^2}{r^2(x^2+y^2)}\nonumber\\
&-&(1-\frac{2m}{r})^{-2\gamma}
\frac{\Delta^{\gamma^2}}{r^2\Sigma^{\gamma^{2}-1}},\\
N^{0033}&=&-\frac{\Delta z^2}{r^4}-\frac{x^2+y^2}{r^2}
-(1-\frac{2m}{r})^{-2\gamma}
\frac{\Delta^{\gamma^2}}{r^2\Sigma^{\gamma^{2}-1}},\\
N^{0012}&=&(\frac{\Delta}{\Sigma})^{\gamma^2-1}\frac{xy}{x^2+y^2}
-\frac{\Delta xy}{r^4}-\frac{xy}{r^2(x^2+y^2)}\nonumber\\
&-&(1-\frac{2m}{r})^{-2\gamma}
\frac{\Delta^{\gamma^2}}{r^2\Sigma^{\gamma^{2}-1}},\\
N^{0013}&=&-\frac{\Delta
xz}{r^4}+\frac{xz}{r^2},\\
N^{0023}&=&-\frac{\Delta yz}{r^4}+\frac{yz}{r^2}.
\end{eqnarray}
Substituting Eqs.(58)-(63) in Eq.(23), we obtain the following
energy density and momentum density components
\begin{eqnarray}
\Omega^{00}&=&\frac{(1-\frac{2m}{r})^{-2\gamma}}{8\pi}[-4\gamma
m^2(2\gamma+1)\frac{\Delta^{\gamma^2-2}}
{r\Sigma^{\gamma^2-1}}+8\gamma m\{\frac{\Delta^{\gamma^2-2}}
{r\Sigma^{\gamma^2-1}}(1-\frac{m}{r})\nonumber\\
&-&(\gamma^2-1)(1-\frac{m}{r})\frac{\Delta^{\gamma^2-1}}
{r\Sigma^{\gamma^2}}-(\frac{\Delta}{\Sigma})^{\gamma^2}\frac{1}{r^3}\}\nonumber\\
&-&2\gamma^2(\gamma^2-1)(1-\frac{m}{r})^2\frac{\Delta^{\gamma^2-2}}
{\Sigma^{\gamma^2-1}}+4\gamma^2(\gamma^2-1)(1-\frac{m}{r})^2
\frac{\Delta^{\gamma^2-1}}{\Sigma^{\gamma^2}}\nonumber\\
&+&(1-\frac{m}{r})\frac{\gamma^2}{r^2}
(\frac{\Delta}{\Sigma})^{\gamma^2-1}-2\gamma^2(\gamma^2-1)
\frac{(x^2+y^2)\Delta^{\gamma^2}}{r^2\Sigma^{\gamma^2+1}}(1-\frac{m}{r}\nonumber\\
&+&\frac{m^2}{r^2}-\frac{m^2(x^2+y^2)}{r^4})^2-2\gamma^2(\gamma^2-1)
\frac{z^2\Delta^{\gamma^2}}{r^2\Sigma^{\gamma^2+1}}
(1-\frac{m}{r}\nonumber\\&-&\frac{m^2(x^2+y^2)}{r^4})^2-
\frac{\gamma^2-1}{r^2}(\frac{\Delta}{\Sigma})^{\gamma^2}
(1-\frac{m}{r}-\frac{2m^2}{r^2}-\frac{3m^2(x^2+y^2)}{r^4})\nonumber\\&-&
\frac{\Delta^{\gamma^2}}{r^4\Sigma^{\gamma^2-1}}]+\frac{1}{8\pi}[
(\gamma^2-1)(1-\frac{m}{r})\{\frac{2x^2}{x^2+y^2}-1\}
\frac{\Delta^{\gamma^2-2}}{\Sigma^{\gamma^2-1}}\nonumber\\
&+&(\gamma^2-1)\frac{\Delta^{\gamma^2-1}}{\Sigma^{\gamma^2}}
(1-\frac{m}{r}+\frac{m^2}{r^2}-\frac{m^2(x^2+y^2)}{r^4})
\{1-\frac{2x^2}{x^2+y^2}\}\nonumber\\
&+&(\frac{\Delta}{\Sigma})^{\gamma^2-1}\{\frac{1}{x^2+y^2}
-\frac{2x^2}{(x^2+y^2)^2}\}-\frac{4}{r^2}(1-\frac{m}{r})+\frac{6\Delta}{r^4}].
\end{eqnarray}

\subsection{Energy and Momentum in M\"{o}ller's Prescription}

For the gamma metric, we obtain the following non-vanishing
components of $K^{bc}_a$
\begin{equation}
K^{01}_0=-2m\gamma sin\theta.
\end{equation}
When we make use of Eq.(65) in Eq.(30), the energy and momentum
density components turn out to be
\begin{equation}
M^0_0=0.
\end{equation}
and
\begin{equation}
M^i_0=0=M^0_i.
\end{equation}
This shows that energy and momentum turn out to be constant.

\section{Conclusion}

Energy-momentum complexes provide the same acceptable
energy-momentum distribution for some systems. However, for some
systems [22-30], these prescriptions disagree. The debate on the
localization of energy-momentum is an interesting and a
controversial problem. According to Misner et al. [35], energy can
only be localized for spherical systems. In a series of papers
[36] Cooperstock et al. has presented a hypothesis which says
that, in a curved spacetime, energy and momentum are confined to
the regions of non-vanishing energy-momentum tensor $T_a^b$ of the
matter and all non-gravitational fields. The results of Xulu
[19,20] and the recent results of Bringley [37] support this
hypothesis. Also, in the recent work, Virbhadra and his
collaborators [14-18] have shown that different energy-momentum
complexes can provide meaningful results. Keeping these points in
mind, we have explored some of the interesting members of the Weyl
class for the energy-momentum distribution.

In this paper, we evaluate energy-momentum densities for the two
solutions of the Weyl metric, i.e., Erez-Rosen and the gamma
metrics. We obtain this target by using four well-known
prescriptions of Einstein, Landau-Lifshitz, Papapetrou and
M$\ddot{o}$ller. From Eqs.(15), (22), (29), (34), (50), (57), (64)
and (67), it can be seen that the energy-momentum densities are
finite and well defined. We also note that the energy density is
different for the four different prescriptions. However, momentum
density components turn out to be zero in all the prescriptions
and consequently we obtain constant momentum for these solutions.
The results of this paper also support the Cooperstock's
hypothesis [36] that energy is localized to the region where the
energy-momentum tensor is non-vanishing.

We would like to mention here that the results of energy-momentum
distribution for different spacetimes are not surprising rather
they justify that different energy-momentum complexes, which are
pseudo-tensors, are not covariant objects. This is in accordance
with the equivalence principle [35] which implies that the
gravitational field cannot be detected at a point. These examples
indicate that the idea of localization does not follow the lines
of pseudo-tensorial construction but instead it follows from the
energy-momentum tensor itself. This supports the well-defined
proposal developed by Cooperstock [36] and verified by many
authors [22-30]. In GR, many energy-momentum expressions
(reference frame dependent pseudo-tensors) have been proposed.
There is no consensus as to which is the best. Hamiltonian's
principle helps to solve this enigma. Each expression has a
geometrically and physically clear significance associated with
the boundary conditions.

\newpage

{\bf Acknowledgment}

\vspace{0.5cm}

We would like to thank for the anonymous referee for his useful
comments.

\vspace{2cm}

{\bf References}

\begin{description}

\item{[1]} Trautman, A.: {\it Gravitation: An Introduction to Current
Research} ed. Witten, L. (Wiley, New York, 1962)169.

\item{[2]} Landau, L.D. and  Lifshitz, E.M.: {\it The Classical Theory of Fields}
(Addison-Wesley Press, 1962).

\item{[3]} Papapetrou, A.: {\it Proc. R. Irish Acad} {\bf A52}(1948)11.

\item{[4]} Tolman R. C: Relativity, Thermodynamics and Cosmology,
(Oxford University Press, Oxford, 1934)227.

\item{[5]} Bergman P.G: and Thompson R. Phys. Rev. {\bf89}(1958)400.

\item{[6]} Weinberg, S.: {\it Gravitation and Cosmology} (Wiley, New York,
1972).

\item{[7]} M\"{o}ller, C.: Ann. Phys. (NY) {\bf 4}(1958)347.

\item{[8]} M\"{o}ller, C.: Ann. Phys. (NY) {\bf12}(1961)118.

\item{[9]} Komar, A. Phys. Rev. {\bf 113}(1959)934.

\item{[10]} Penrose, R.: {\it Proc. Roy. Soc.} London {\bf
A388}(1982)457;\\
{\it GR10 Conference}, eds. Bertotti, B., de Felice, F. and
Pascolini, A. Padova {\bf 1} (1983)607.

\item{[11]} Brown, J.D. and York, J,W.: Phys. Rev. {\bf D47}(1993)1407.

\item{[12]} Hayward, S.A.: Phys. Rev. {\bf
D49}(1994)831.

\item{[13]} Chang, C.C., Nester, J.M. and Chen, C.: Phys. Rev. Lett. {\bf
83}(1999)1897.
\item{[14]} Virbhadra, K.S.: Phys. Rev. {\bf D42}(1990)2919.

\item{[15]} Virbhadra, K.S.: Phys. Rev. {\bf D60}(1999)104041.

\item{[16]} Rosen, N. and Virbhadra, K.S.: Gen. Relati. Gravi. {\bf 25}(1993)429.

\item{[17]} Virbhadra, K.S. and Parikh, J.C.: Phys. Lett. {\bf B317}(1993)312.

\item{[18]} Virbhadra, K.S. and Parikh, J.C.: Phys. Lett. {\bf B331}(1994)302.

\item{[19]} Xulu, S.S.: Int. J. of Mod. Phys. {\bf A15}(2000)2979;
Mod. Phys. Lett. {\bf A15}(2000)1151 and reference therein.

\item{[20]} Xulu, S.S.: Astrophys. Space Sci. {\bf 283}(2003)23.

\item{[21]} Aguirregabiria, J.M., Chamorro, A. and Virbhadra, K.S,:
{\it Gen. Relativ. and Grav.} {\bf 17}, 927 {\bf 28}(1996)1393.

\item{[22]} Sharif, M.: Int. J. of Mod. Phys. {\bf A17}(2002)1175.

\item{[23]} Sharif, M.: Int. J. of Mod. Phys. {\bf A18}(2003)4361;
Errata {\bf A19}(2004)1495.

\item{[24]} Sharif, M.: Int. J. of Mod. Phys. {\bf D13}(2004)1019.

\item{[25]} Gad, R.M: Astrophys. Space Sci. {\bf 293}(2004)453.

\item{[26]} Gad, R.M: Mod. Phys. Lett. {\bf A19}(2004)1847.

\item{[27]} Patashnick, O.: Int. J. of Mod. Phys. {\bf D}(2005) (gr-qc/0408086).

\item{[28]} Sharif, M. and Fatima, Tasnim: Int. J. of Mod. Phys. {\bf A20}(2005)4309.

\item{[29]} Sharif, M. and Fatima, Tasnim: Nuovo Cimento {\bf B}(2005).

\item{[30]} Fatima, Tasnim: M.Phil. Thesis (University of the Punjab, Lahore, 2004).

\item{[31]} Weyl, H.: Ann. Phys. (Leipzig) {\bf 54}(1917)117;
{\bf 59}(1919)185;\\
Civita, Levi, L.: Atti. Acad. Naz. Lince Rend. Classe Sci. Fis.
Mat. e Nat., {\bf 28}(1919)101;\\
Synge, J.L.: {\it Relativity, the General Theory} (North-Holland
Pub. Co. Amsterdam, 1960).

\item{[32]} Kramer, D., Stephani, H., MacCallum, M.A.H. and Hearlt, E.: {\it
Exact Solutions of Einstein's Field Equations} (Cambridge
University Press, 2003).

\item{[33]} Curzon, H.E.J.: {\it Proc. Math. Soc.} London {\bf 23}(1924)477.

\item{[34]} Esposito, F. and Witten, L.: Phys. Lett. {\bf
B58}(1975)357;\\
Virbhadra, K.S.: gr-qc/9606004;\\
Herrera, L., A Di Prisco, A.Di. and Fuenmayor, E.: Class. Quant.
Grav. {\bf 20}(2003)1125.

\item{[35]} Misner,C.W., Thorne, K.S. and Wheeler, J.A. {\it
Gravitation} (W.H. Freeman, New York, 1973)603.

\item{[36]} Cooperstock, F.I. and Sarracino, R.S. {\it J. Phys. A.:
Math. Gen.} {\bf 11}(1978)877. \\
Cooperstock, F.I.: in {\it Topics on Quantum Gravity and Beyond},
Essays in honour of Witten, L. on his retirement, ed. Mansouri, F.
and Scanio, J.J. (World Scientific, Singapore, 1993); Mod. Phys.
Lett. {\bf A14}(1999)1531; Annals of Phys. {\bf
282}(2000)115;\\
Cooperstock, F.I. and Tieu, S.: Found. Phys. {\bf 33}(2003)1033.

\item{[37]} Bringley, T.: Mod. Phys. Lett. {\bf A17}(2002)157.

\end{description}

\end{document}